\documentclass[12pt]{article}
\usepackage{amsmath,amsfonts,amssymb,mathtools}
\usepackage[nosort]{cite}
\usepackage{hyperref}
\usepackage{graphicx}
\usepackage{color}
\usepackage{epsfig}
\usepackage{psfrag}	
\usepackage{tikz}
\usetikzlibrary{arrows.meta,shapes,chains}
\usepackage{slashed}
\usepackage{tensor}
\usepackage{ulem}
\usepackage{bbm}
\usepackage{fancyref}
\usepackage[font=small,labelfont=bf,width=1\textwidth]{caption}
\usepackage{comment}
\usepackage[all]{xy}
\usepackage{multirow}
\usepackage{float}
\usepackage{cite}
\usepackage{color}
\usepackage{xcolor}

\usepackage{pifont}
\usepackage{tensor}

\usepackage{tocloft} 
\usepackage{ulem}
\usepackage{soul,xcolor}
\setstcolor{red}

\numberwithin{equation}{section}

\DeclareMathOperator{\sTr}{sTr}

\DeclareMathOperator{\diag}{diag}

\newcommand{\bea}{\begin{eqnarray}}
\newcommand{\eea}{\end{eqnarray}}
\newcommand{\beq}{\begin{equation}}
\newcommand{\eeq}{\end{equation}}
\newcommand{\bal}{\begin{equation}\begin{aligned}}
\newcommand{\eal}{\end{aligned} \end{equation}}

\newcommand{\address}[1]{\vbox{\center\em#1}}
\renewcommand{\title}[1]{\vbox{\center\huge{#1}}\vspace{5mm}}

\newcommand{\cA}{{\mathcal A}}

\newcommand{\cD}{{\mathcal D}}

\newcommand{\cL}{{\mathcal L}}

\newcommand{\cN}{{\mathcal N}}
\newcommand{\cP}{{\mathcal P}}
\newcommand{\cQ}{{\mathcal Q}}

\newtheorem{algorithm}{Algorithm}

\begin{document}
\begin{titlepage}
\begin{center}

\vspace*{20mm}

\title{A network of hyperloops}

\vspace{7mm}

\renewcommand{\thefootnote}{$\alph{footnote}$}

Ziwen Kong%
\footnote{\href{mailto:ziwen.kong@kcl.ac.uk}
{\tt ziwen.kong@kcl.ac.uk}}

\vskip 2mm
\address{
Department of Mathematics, King's College London,
\\
The Strand, WC2R 2LS London, United-Kingdom}

\renewcommand{\thefootnote}{\arabic{footnote}}
\setcounter{footnote}{0}

\end{center}
\vspace{4mm}
\abstract{
\normalsize{
\noindent
In this paper we complete the exploration of connected components of the space of BPS Wilson loops in three-dimensional $\cN=4$ Chern-Simons-matter theory on $S^3$. The algorithm is to start with a supersymmetric Wilson loop, choose a preserved supercharge, and look for BPS deformations built out of the matter ﬁelds in the proper representations. Using this, we discover many new moduli spaces of nonconformal BPS Wilson loops preserving a single or two supercharges, which are subsets of the symmetries of the 1/4 and 3/8 BPS operators. Along with the those previously found in \cite{Drukker:2020dvr,Drukker:2022ywj} and \cite{Drukker:2022bff}, the total moduli spaces are closed under this formalism.
}}
\vfill

\end{titlepage}

\tableofcontents

\section{Introduction and conclusions}
In three-dimensional supersymmetric conformal ﬁeld theories there are vast moduli spaces of BPS Wilson loops. In addition to the bosonic loops that couple to only one gauge ﬁeld and bilinear of the scalars \cite{Gaiotto:2007qi}, most BPS Wilson loops have the superconnections $\cL$ comprised of at least two vector ﬁelds as well as the matter ﬁelds in figure \ref{fig:N=4quiver}
\bal
W= \sTr \cP \exp i \oint \cL d\varphi\,.
\eal
More and more such examples have been found in the past years \cite{Drukker:2020dvr, Drukker:2022ywj, Drukker:2022bff, Rey:2008bh, Drukker:2008zx, Chen:2008bp, Cooke:2015ila, Ouyang:2015qma, Ouyang:2015iza, Ouyang:2015bmy, Mauri:2017whf, Mauri:2018fsf, Drukker:2019bev, Drukker:2020opf,  Castiglioni:2022yes}. For a recent review about what was known at the time on this topic, see \cite{Drukker:2019bev}.

To discover new BPS Wilson loops, instead of relying on complicated ansatze, there is a more efficient algorithm constructed in the series of ``Hyperloop" papers \cite{Drukker:2020dvr, Drukker:2022ywj, Drukker:2022bff}
\begin{algorithm}\leavevmode
\label{itm:algorithm}
\begin{enumerate}
    \item[1.] Pick a BPS Wilson loop of the theory with (super)connection $\cL$.
    \item[2.] Choose a supercharge it preserves, $\cQ$.
    \item[3.] Look for deformations that still preserve that supercharge.
\end{enumerate}
\end{algorithm}
The hyperloops refer to the supersymmetric Wilson loops, especially those coupling to the fermionic fields in $\cN = 4$ supersymmetric Chern-Simons-matter theories with either linear or circular quiver structure \cite{Gaiotto:2008sd, Imamura:2008dt, Hosomichi:2008jd, Hama:2010av, Hama:2011ea} on $S^3$ along a great circle. However, this algorithm can be easily applied to other supersymmtric theories such as ABJM\footnote{Actually this algorithm is firstly proposed by \cite{Drukker:2019bev} in ABJM theory.}  Step 3 of this algorithm centers on the formula
\bal
\label{generaldef1}
\cL \rightarrow \cL-i \cQ G+ \{H,G\} +\Pi \tilde{\Pi} G^2+C\, ,
\eal
where $\Pi, \tilde{\Pi}$ \eqref{Pi}, \eqref{tildePi}are parameters related to the supercharge $\cQ$, and $H$ is determined by the relaxed supersymmetry condition
\bal
\label{susycon}
\cQ \cL= \mathcal{D}_{\varphi}^{\cL} H \, ,
\eal
and $G,C$ are some other supermatrices that will be explained in detail later  \eqref{G1/4}, \eqref{G01/4}, \eqref{C1/4}.

The first Hyperloop paper \cite{Drukker:2020dvr} studied the cases with bosonic loops as the starting points, and deformed with certain linear combinations of supercharge $\cQ$. The resulting moduli spaces are composed of hyperloops preserving the same supercharge.
The second Hyperloop paper \cite{Drukker:2022ywj} focused on the deformations of 1/2 BPS loops with arbitrary linear combination of the supercharges. Since there are eight supercharges preserved by a 1/2 BPS loop, in this way, we discovered an eight-dimensional space, where each point corresponds to one moduli space generated by a fixed supercharge. These moduli spaces intersect where hyperloops preserve more than one supercharge, for example loops with $SU(2)$ R-symmetry enhancements.

We dub the connected components of moduli spaces as the ``network". In this paper, instead of a fixed starting point, we allow $\cL$ to travel along the network, and look for all possible moduli spaces produced by the algorithm. At each step of the itinerary we run the whole algorithm \ref{itm:algorithm} to obtain the consequent moduli spaces with \eqref{generaldef1}, and the starting point $\cL$ and the supercharge\footnote{In this paper, we do not consider $\cQ$ combined with only supercharges in the same chirality.}  $\cQ$ are chosen as below
\begin{itemize}
    \item[i)] Starting with the 1/2 BPS loops and arbitrary non-nilpotent supercharges\footnote{The $\Pi\ne 0$ supercharges in \cite{Drukker:2022ywj}.} preserved by it, the resulting moduli spaces are found in \cite{Drukker:2022ywj}, 
    \item[ii)] Starting with the bosonic loops\footnote{There are some subtleties that we will explain later around \eqref{bos}.} \cite{Drukker:2022bff} and arbitrary supercharges preserved by them. In the resulting moduli spaces, there are some special points that receive supersymmtry enhancements, especially those preserving $SU(2)$ R-symmetry (thus being 1/2, 3/8 and 1/4 BPS).
    \item[iii)] Starting with the $SU(2)$ enhanced points and any preserved supercharge including both nilpotent and non-nilpotent ones.
\end{itemize}

Note that in step ii) we actually employ a further trick. The 1/2 BPS loops we start with are built around the adjacent $I$ and $I$+1 nodes, when it comes to the bosonic loops, since they contain no terms that are linear in the matter fields, one may decouple the nodes and rebuild a superconnection around the $I$+1 and $I$+2 ones, which are the starting point of the second deformation. All of the new hyperloops constructed here are in this setting. Alternatively, one can also take the superconnections in larger supermatrices that are built around $I$, $I$+1, $I$+2 nodes and study BPS deformations in such cases, like in the section 5 of \cite{Drukker:2022ywj}.

\begin{figure}[h]
    \centering
    \tikzstyle{format}=[rectangle,draw,thin,fill=white]
    \tikzstyle{test}=[diamond,aspect=2,draw,thin]
    \tikzstyle{point}=[coordinate,on grid,]
    \begin{tikzpicture}
        \node[format](start1) at (0,0){\cite{Drukker:2020dvr,Drukker:2022ywj} Start};
        \node[format,below of=start1,node distance=10mm](WL1){Choose $\cL$};
        \node[format,below of=WL1,node distance=10mm](supercharge1){Choose $\cQ$};
        \node[format,below of=supercharge1,node distance=10mm](deformation1){Deformation \eqref{generaldef1}};
        \node[format,below of=deformation1,node distance=10mm](publish1){Publish!};

        \node[format](start) at (6,0){Start};
        \node[format,below of=start,node distance=10mm](WL){Choose $\cL$};
        \node[format,below of=WL,node distance=10mm](supercharge){Choose $\cQ$};
        \node[format,below of=supercharge,node distance=10mm](deformation){Deformation \eqref{generaldef1}};
        \node[test,below of=deformation,node distance=16mm](test){New points?};
        \node[format,below of=test,node distance=18mm](publish){Publish!};

        \node[point,right of=test,node distance=25mm](point1){};
        \node[point,right of=WL,node distance=25mm](point2){};

        \draw[-Latex](start1)--(WL1);
        \draw[-Latex](WL1)--(supercharge1);
        \draw[-Latex](supercharge1)--(deformation1);
        \draw[-Latex](deformation1)--(publish1);

        \draw[-Latex](start)--(WL);
        \draw[-Latex](WL)--(supercharge);
        \draw[-Latex](supercharge)--(deformation);
        \draw[-Latex](deformation)--(test);
        \draw[-](test)--node[below]{Yes}(point1);
        \draw[-](point1)--(point2);
        \draw[-Latex](point2)--(WL);
        \draw[-Latex](test)--node[right]{No}(publish);
    \end{tikzpicture}
    \caption{Strategies used in \cite{Drukker:2020dvr,Drukker:2022ywj} and this paper. ``New points" in the conditional block refers to those with supersymmetry enhancements.}
    \label{fig:flowchart}
\end{figure}
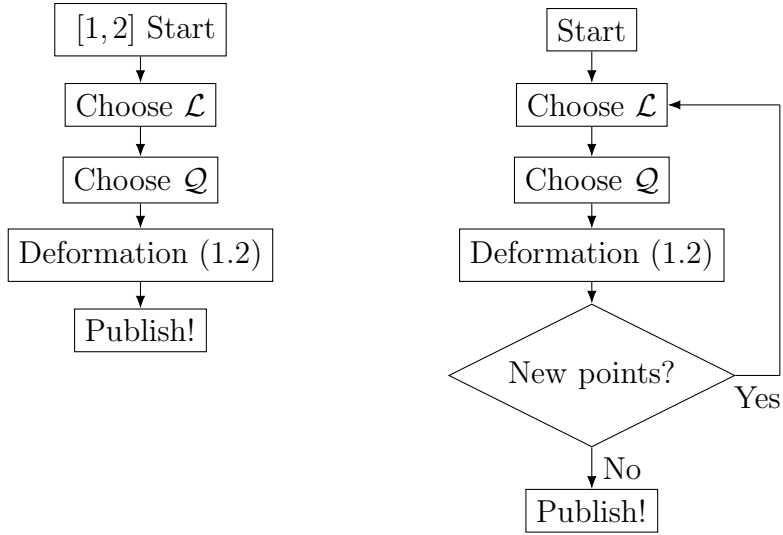

In fact, we can choose any supersymmetric loop in the network as the new starting point for the deformation, which will always take the same form as \eqref{generaldef1} with the replacement of the corresponding $\Pi^{\prime}, \tilde{\Pi}^{\prime}$ and $\cQ^{\prime}, H^{\prime}$. However, the resulting moduli spaces will be subspaces of those generated in i), ii) and iii). In other words, the network is closed under the algorithm. Compared to the previous Hyperloop papers, the difference in the implementation of algorithm \ref{itm:algorithm} is summarised in the flowcharts \ref{fig:flowchart}

Let us review some of the resulting network. One special kind of loops is the bosonic loops. Their full classification in $\cN=4$ Chern-Simons matter theory is given in the third Hyperloop paper \cite{Drukker:2022bff}, where in the notation employed here, all of them can be summarized as\footnote{Except for two special cases given in section 3.3.3 and 3.4 in \cite{Drukker:2022bff} where the preserved supercharges are composed entirely of either barred or unbarred ones.}
\bal
\label{Lbos}
\cA^{\mathrm{bos}}= \cA_{\varphi}+\frac{i}{k\Pi} (r^1 \bar{r}_1 -r^2 \bar{r}_2) -\frac{i}{k \tilde{\Pi}} (\bar{\tilde{r}}^{\dot{1}} \tilde{r}_{\dot{1}} -\bar{\tilde{r}}^{\dot{2}} \tilde{r}_{\dot{2}})\,,
\eal
where $r,\tilde{r}$ \eqref{rotatedscalars}, \eqref{rotatedchi} are the rotated scalar fields. Because of the full classification of bosonic loops, in step ii) we exhaust all the hyperloops \eqref{Lfer} with
\bal
\label{su2enhancement}
\Tilde{\cA}=A_{\varphi}+\frac{i}{k\Pi}(r^1 \bar{r}_1 -r^2\bar{r}_2)+\frac{i}{k} \Tilde{\nu}
\eal
in the diagonal entries. By setting them as the new starting point in step iii) and perform the supersymmetric deformations, we find all the moduli spaces of hyperloops that are connected to theirs. The expressions \eqref{su2enhancement} are of the hyperloops that are built around $I$+1 and $I$+2 nodes, of course the same formalism can be applied to any other two adjacent nodes.

All the hyperloops in this network can be classified into two types. One is what we call $\Pi \ne 0 $ (or $\tilde{\Pi} \ne 0)$ loops that can be deformed from all the possible bosonic loops and arbitrary preserved supercharge $\cQ$. The other one is what we call $\Pi= 0$ (or $\tilde{\Pi}=0$)\footnote{$\Pi$ and $\tilde{\Pi}$ do not vanish at the same time.} loops, first found in \cite{Drukker:2022ywj}, where the preserved supercharges $\cQ$ are nilpotent and some scalars are annihilated by them. They are produced in step i) and iii) and do not live in the moduli spaces generated by the deformation from bosonic loops ii).  In particular, in this case we find other loops that preserves $SU(2)_L$ (or $SU(2)_R$) R-symmetries.

Then a question naturally comes up, whether these moduli spaces are really complete? Although we have exhausted all the connected moduli spaces of hyperloops through our construction, we are still unable to give a definite answer since there might be some isolated components in the complete moduli space. It would be very interesting to look for such BPS Wilson loops. In particular, a longstanding question was whether 1/3 BPS Wilson loops exist ABJM theory. This was answered in the recent paper \cite{Drukker:2022txy}, but when we set the 1/3 BPS loop as the starting point and played our algorithm, we did not find more 1/3 BPS loops up to global gauge symmetries. Therefore, if other 1/3 BPS loops really exist, they probably live in the isolated points.

Another question is about Wilson loops with superconnections in larger supermatrices, especially those coupling to repetitive gauge fields \cite{Drukker:2020opf}. We attempted to set the starting point as two copies of 1/2 BPS loops \eqref{half1}, in which case the gauge symmetry is $S(GL(2,\mathbb{C})\times GL(2,\mathbb{C}))$, where $S$ denotes that the center of $GL(4,\mathbb{C})$ is excluded. The outcome is very similar to \cite{Drukker:2022ywj} and does not include new conformal hyperloops in the moduli spaces, expect for the ones that are direct sums of two conformal loops in $2\times2$ supermatrices \cite{Drukker:2022ywj} up to the gauge transformation, so we skip the presentation of this in the paper.

Other future directions are very similar to those mentioned in \cite{Drukker:2022ywj}. One is to compute the expectation value of all the hyperloops in the network, among which cases of Gaitto-Yin loops and their fermionic deformations are studied in \cite{Cooke:2015ila,Drukker:2022ywj,Drukker:2008jm} and the ``latitude" loops are in \cite{Griguolo:2015swa,Bianchi:2016vvm, Bianchi:2016yzj,Bianchi:2018bke}. Another direction is about the holographic duals of hyperloops \cite{Ouyang:2015qma, Cooke:2015ila, Lietti:2017gtc, Correa:2019rdk, Correa:2021sky, Bianchi:2020hsz, Garay:2022szq}. Besides, since all the hyperloops newly found in this paper are not conformal, they do not contribute to the defect conformal manifolds \cite{Callan:1994ub, Recknagel:1998ih, Gaberdiel:2008fn,Behan:2017mwi, Karch:2018uft, Drukker:2022pxk}. However, it would be interesting to understand their renormalization group ﬂows \cite{Polchinski:2011im, Cuomo:2021rkm}.

This paper is organised as follows. In the next section we present notations for the theories and supersymmetry variations of the ﬁelds, as well as details of the bosonic loops \eqref{Lbos} and the hyperloops with $SU(2)_L$ and $SU(2)_R$ supersymmetry enhancements, which by the amount of preserved supercharges can be classified into 1/4, 3/8 and 1/2 BPS loops. In Sections \ref{1/4 BPS Deformation} and \ref{3/8 BPS loops} we discuss the moduli spaces produced by deformations from 1/4 and 3/8 BPS hyperloops\footnote{The deformation from 1/2 BPS hyperloops have been studied in \cite{Drukker:2022ywj}.} respectively. The supersymmetry transformation rules are collected in appendix.

\section{${\mathcal{N}}=4$ Chern-Simons-matter theories on unit $S^3$ and hyperloops}
\label{sec:notation}

\begin{figure}[H]
\centering
\begin{tikzpicture}
\draw[line width=.5mm] (6,2) circle (7mm);
\draw[line width=.5mm] (10,2) circle (7mm);
\draw[line width=.5mm] (2.7,2)--(5.3,2);
\draw[line width=.5mm] (6.7,2)--(9.3,2);
\draw[line width=.5mm] (10.7,2)--(13,2);
\draw (6,2) node [] {$A_{I}$};
\draw (10,2) node [] {$A_{I+1}$};
\draw (6,1) node [] {$-k$};
\draw (10,1) node [] {$k$};
\draw (4.05,2.4) node [] {$\tilde{q}_{I-1\,\dot{a}}, \ \tilde{\psi}^a_{I-1}$};
\draw (4.05,1.6) node [] {$\bar{\tilde{q}}^{\dot{a}}_{I-1}, \bar{\tilde{\psi}}_{I-1\,a}$};
\draw (8,2.4) node [] {$\bar{q}_{I\,a}, \ \bar\psi^{\dot{a}}_{I}$};
\draw (8,1.6) node [] {$q^a_{I},\ \psi_{I\,\dot{a}}$};
\draw (12,2.4) node [] {$\tilde{q}_{I+1\,\dot{a}},\ \tilde{\psi}^a_{I+1}$};
\draw (12,1.6) node [] {$\bar{\tilde{q}}^{\dot{a}}_{I+1},\ \bar{\tilde{\psi}}_{I+1\,a}$};
\end{tikzpicture}
\caption{The quiver and field content of the ${\cal N}=4$ theory.}
\label{fig:N=4quiver}
\end{figure}
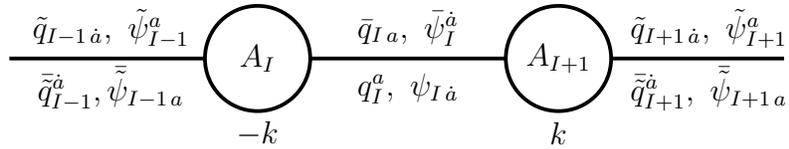

Our setting is $\cN=4$ Chern-Simons-matter theories, with either a circular or a linear quiver diagram \cite{Gaiotto:2008sd, Imamura:2008dt, Hosomichi:2008jd, Hama:2010av}. In the quiver diagram \ref{fig:N=4quiver}, the edges represent hypermultiplets $(q_I^a,\psi_{I\dot a})$ and the twisted hypermultiplet $(\tilde q_{I-1\,\dot a},\tilde\psi_{I-1}^{a})$, and so on in an alternate fashion.

The fields with indices $a,b=1,2$ are doublets of the $SU(2)_L$ R-symmetry, while those with indices $\dot a,\dot b=\dot 1,\dot 2$ are of $SU(2)_R$ R-symmetry. These indices are raised and lowered using the appropriate epsilon 
symbols: $v^a=\epsilon^{ab}v_ b$ and $v_a=\epsilon_{ab}v^b$ with $\epsilon^{12}=\epsilon_{21}=1$, 
and similarly for the dotted indices.

To write down the hyperloops and the supersymmetry variations in appendix \ref{susytrans}, following \cite{Cooke:2015ila, Drukker:2020dvr} it is useful to define
\bal
\mu_{I}{}^a_{\ b}&=q_I^a\bar q_{I\,b}-\frac{1}{2}\delta^a_bq_I^c\bar q_{I\,c}\,,
\qquad& \nu_{I}&=q_I^a\bar q_{I\,a}
\,,\\
\tilde\mu_{I}{}^{\dot a}_{\ \dot b}&=\bar{\tilde q}_{I-1}^{\,\dot a}\tilde q_{I-1\,\dot b}
-\frac{1}{2}\delta^{\dot a}_{\dot b}\bar{\tilde q}_{I-1}^{\,\dot c}\tilde q_{I-1\,\dot c}\,,
\qquad&
\tilde\nu_{I}&=\bar{\tilde q}_{I-1}^{\,\dot a}\tilde q_{I-1\,\dot a}\,.
\eal
These are bilinears of the matter fields and transform in the 
adjoint representation of $U(N_I)$. In the same way, we can construct other bilinears 
that transform in the adjoint of other $U(N)$'s. 
The supersymmetry transformations are given in appendix \ref{susytrans}.

Our first example of a hyperloop in the theory is the 1/2 BPS loop \cite{Drukker:2009hy}. We take a 1/2 BPS Wilson loop built around the $I$ and $I+1$ nodes, whose superconnection is in the $GL(N_I|N_{I+1})$ supermatrix preserving $SU(2)_L$ symmetry
\bal
\label{half1}
\cL_{I,1/2}=\begin{pmatrix}
    \cA_I & -i \bar{\alpha} \psi_{I \dot{1}-}\\
    i \alpha \bar{\psi}_{I+}^{\dot{1}} & \cA_{I+1}-\frac{1}{2}
\end{pmatrix}\, ,
\eal
where
\bal
\cA_{I}=A_{\varphi,I}+\frac{i}{k}(\nu_I- \tilde\mu_{I}{}^{\dot 1}_{\ \dot 1}+ \tilde\mu_{I}{}^{\dot 2}_{\ \dot 2}),\quad \cA_{I+1}=A_{\varphi,I+1}+\frac{i}{k}(\nu_{I+1}- \Tilde{\mu}_{I+1\dot{1}}{}^{\dot{1}}+ \Tilde{\mu}_{I+1\dot{2}}{}^{\dot{2}})\, ,
\eal
and the constants $\alpha$ and $\bar{\alpha}$ (which are not complex conjugate to each other) satisfy $\alpha \bar{\alpha} =2i/k$. The Wilson loop does not depend on their actual value, since the loops are invariant under the (constant) gauge transformation \cite{Drukker:2020opf}
\bal
\cL_{I,1/2} \rightarrow \begin{pmatrix}
    1 & 0\\
    0 & 1/x
\end{pmatrix} \cL_{I,1/2} \begin{pmatrix}
    1 & 0\\
    0 & x
\end{pmatrix}\, ,
\eal
the equivalence relation $(\bar{\alpha}, \alpha) \sim (\bar{\alpha} x, \alpha/x)$ gives the global gauge symmetry $\mathbb{C}^{\star}$, so the moduli space is just one point. We could also allow for $\alpha,\bar{\alpha}$ to depend on $\varphi$ at the expense of a $U(1)$ gauge transformation at the bottom right entry
\bal
\cA_{I+1}-\frac{1}{2} \rightarrow \cA_{I+1}- \frac{1}{2} -i \alpha ^{-1} \partial_{\varphi} \alpha\,.
\eal
This shift in the connection can be compared with another 1/2 BPS loop, which includes the same gauge ﬁelds and preserves the exact same symmetries \cite{Cooke:2015ila}, but couples instead to other ﬁelds in the hypermultiplets
\bal
\label{half2}
\cL_{I,1/2}^{\prime} =\begin{pmatrix}
    \cA_I^{\prime} & i \bar{\alpha} \psi_{I \dot{2}+}\\
    i\alpha \bar{\psi}_{I-}^{\dot{2}} & \cA_{I+1}^{\prime} +\frac{1}{2}
\end{pmatrix}\, ,
\eal
where $\cA_{I}^{\prime} =A_{\varphi,I}-\frac{i}{k}(\nu_I +\Tilde{\mu}_{I\;\dot{1}}^{\; \dot{1}}- \Tilde{\mu}_{I\;\dot{2}}^{\; \dot{2}})\,$. The shift $+1/2$ can be transformed to $-1/2$ by bringing in extra phases to $\bar{\alpha}^{\prime}, \alpha^{\prime}$. These two loops preserve the same eight supercharges, where we can write a general superposition of them as
\bal
\label{Q1/2}
\cQ_{1/2}=\eta_a^{\imath} Q_{\imath}^{\dot{2}a+} +\bar{\eta}_a^{\imath} (\sigma^1)_{\imath}^{\bar{\imath}} Q_{\bar{\imath}}^{\dot{1}a-}\, .
\eal

Analogous to \cite{Drukker:2022ywj}, we start with either of the two 1/2 BPS loops and play the algorithm with \eqref{Q1/2}. Among all the resulting moduli spaces, in particular we pay our attention to the bosonic loops where the nodes get decoupled
\bal
\label{bos}
\cA^{\mathrm{bos}}=A_{\varphi} +\frac{i}{k \Pi} (r^1 \bar{r}_1 -r^2 \bar{r}_2) -\frac{i}{k} (\tilde{\mu}_{\;\dot{1}}^{\dot{1}} - \tilde{\mu}_{\;\dot{2}}^{\dot{2}})\, ,
\eal
with
\bal
\label{Pi}
\Pi\equiv \epsilon^{ab} (\bar{\eta} v)_a (\eta \bar{v})_b\, .
\eal
The rotated scalar fields $r,\bar{r}$ are defined as
\bal
\label{rotatedscalars}
r^1\equiv ( \eta \bar v)_a q^a \,,\quad
r^2\equiv ( \bar\eta v)_a q^a\,,&\quad
\bar r_1\equiv \epsilon^{ab}(\bar\eta v)_a \bar q_b \,,\quad
\bar r_2\equiv -\epsilon^{ab}(\eta \bar v)_a \bar q_b\, ,
\eal
where $(\eta\bar v)_a=\eta^\imath_a \bar v_\imath$ and likewise for $(\bar\eta v)_a$, with the auxiliary vectors
\bal
v_{\imath}=\begin{pmatrix}
e^{+i\varphi}\\
1
\end{pmatrix}_{\imath},\quad \bar{v}_{\imath}=\begin{pmatrix}
1\\
e^{-i\varphi}
\end{pmatrix}_{\imath}\,.
\eal

The bosonic loops \eqref{bos} preserve (at least) two supercharges $\eta_a^{\imath} Q_{\imath}^{\dot{2}a+}$ and $\bar{\eta}_a^{\imath} (\sigma_1)_{\imath}^{\bar{\imath}} Q_{\bar{\imath}}^{\dot{1}a-}$. As discussed in \cite{Drukker:2022bff}, we can further classify them into the following cases:
\begin{itemize}
\item When both $\eta_a^{\imath}$ and $\bar{\eta}_a^{\imath}$ factorise, i.e. $\eta_a^{\imath}=y^{\imath} w_a$, $\bar{\eta}_a^{\imath}=\bar{y}^{\imath} \bar{w}_a$, \eqref{bos} preserves 4 supercharges and thus is 1/4 BPS. In particular, with the choice $w_a=\delta_a^2$ and $\bar{w}_a=\delta_a^1$, we recover the Gaiotto-Yin loop \cite{Gaiotto:2007qi}
\bal
\cA_{\mathrm{GY}}^{\mathrm{bos}}=A_{\varphi} -\frac{i}{k} (q^1 \bar{q}_1- q^2 \bar{q}_2)- \frac{i}{k} (\bar{\tilde{q}}^{\dot{1}} \tilde{q}_{\dot{1}} -\bar{\tilde{q}}^{\dot{2}} \tilde{q}_{\dot{2}})\,.
\eal

\item When one of $\eta_a^{\imath},\bar{\eta}_a^{\imath}$ factorises while the other does not, we get a 3/16 BPS bosonic loop \cite{Drukker:2022ywj,Drukker:2022bff}. For instance, when $\bar{\eta}_1^l =\bar{\eta}_2^r =\eta_2^l=1$ and other $\eta,\bar{\eta}$'s vanish, it leads to the loop
\bal
\cA_{\mathrm{3/16}}^{\mathrm{bos}}=A_{\varphi} -\frac{i}{k} (q^1 \bar{q}_1+2e^{-i\varphi} q^2 \bar{q}_1 - q^2 \bar{q}_2)- \frac{i}{k} (\bar{\tilde{q}}^{\dot{1}} \tilde{q}_{\dot{1}} -\bar{\tilde{q}}^{\dot{2}} \tilde{q}_{\dot{2}})\,.
\eal

\item When neither $\eta_a^{\imath}$ nor $\bar{\eta}_a^{\imath}$ factorises, the resulting loop is 1/8 BPS. For example,
when 
\bal
\label{lattitudeparameter}
\bar{\eta}_1^r=\eta_2^l= \cos \frac{\theta}{2},\quad \bar{\eta}_2^l=-\eta_1^r= \sin \frac{\theta}{2}\, ,
\eal
and other $\eta,\bar{\eta}$'s vanish, the resulting loops are the “bosonic latitude loops \cite{Hama:2010av, Drukker:2020dvr, Cardinali:2012ru}
    \bal
    \cA_{\mathrm{LA}}^{\mathrm{bos}}=A_{\varphi} +\frac{i}{k} M_{LA} q^a \bar{q}_b -\frac{i}{k}(\tilde{\mu}_{\;\dot{1}}^{\dot{1}}- \tilde{\mu}_{\;\dot{2}}^{\dot{2}}),\quad  M_{LA}=\begin{pmatrix}
        \cos \theta & \sin \theta e^{-i\varphi}\\
        \sin \theta e^{i\varphi} & -\cos \theta
    \end{pmatrix}\, .
    \eal   
And when
\bal
\label{Jordannormal}
\begin{pmatrix}
    \bar{\eta}_1^l & \bar{\eta}_1^r\\
    \bar{\eta}_2^l & \bar{\eta}_2^r
\end{pmatrix}=\begin{pmatrix}
    1 & 0\\
    0 & 1
\end{pmatrix},\quad \begin{pmatrix}
    \eta_1^l & \eta_1^r\\
    \eta_2^l & \eta_2^r
\end{pmatrix}=\begin{pmatrix}
    1 & 1\\
    0 & 1
\end{pmatrix},
\eal
the resulting loops are the non-trivial Jordan normal form bosonic loops in section 3.3.1 of \cite{Drukker:2022bff}
    \bal
    \label{JNbosloops}
    \cA_{\mathrm{JN}}^{\mathrm{bos}}=A_{\varphi} +\frac{i}{k} M_{JN} q^a \bar{q}_b -\frac{i}{k}(\tilde{\mu}_{\;\dot{1}}^{\dot{1}}- \tilde{\mu}_{\;\dot{2}}^{\dot{2}}),\;\;  M_{JN}=\begin{pmatrix}
        1+2e^{i \varphi} & -2 (e^{i \varphi} +e^{2 i \varphi})\\
        2 & -(1+2e^{i \varphi})
    \end{pmatrix} .
    \eal
\end{itemize}

Note that not all the bosonic loops found in \cite{Drukker:2022bff} can be represented as \eqref{bos}. There are some other 1/8 BPS loops in section 3.3.2 of \cite{Drukker:2022bff} that will be presented later in \eqref{bos3/16?}.

Furthermore, from the deformation of the bosonic loops \eqref{bos} we can get the general fermionic hyperloops in $GL(U_{I+1}|U_{I+2})$ supermatrices that preserves $SU(2)_R$ symmetry
\bal
\label{Lfer}
\mathcal{L}=\begin{pmatrix}
\Tilde{\cA}_{I+1}-\frac{1}{2} & -i \bar{\delta}^{\dot{1}} \Tilde{\rho}_+^2\\
-i \frac{\delta_{\dot{1}}}{\Pi} \bar{\Tilde{\rho}}_{2-} & \Tilde{\cA}_{I+2}+\bar{\Gamma}
\end{pmatrix}\, ,
\eal
where $\Pi \ne 0$ and
\bal
\label{cAfer}
\Tilde{\cA}=A_{\varphi}+\frac{i}{k\Pi}(r^1 \bar{r}_1 -r^2\bar{r}_2)+\frac{i}{k} \Tilde{\nu},\quad \bar{\Gamma}=\frac{1}{2}\left(i \partial_{\varphi} \log \Pi -\frac{\lambda}{\Pi} -1\right)\, ,
\eal
with 
\bal
\label{lambda}
\lambda=\epsilon^{ab} \epsilon_{\imath \jmath} \bar{\eta}_a^{\imath} \eta_b^{\jmath}
\eal
satisfying
\bal
\label{lineareqn}
\epsilon^{ab} (\bar{\eta} v)_a (\eta \sigma^3 \bar{v})_b =-i\partial_{\varphi} \Pi -\lambda,\quad \epsilon^{ab} (\bar{\eta} \sigma^3 v)_a (\eta \bar{v})_b=-i\partial_{\varphi} \Pi +\lambda\,.
\eal
The rotated fermions $\bar{\rho}_+^2, \bar{\tilde{\rho}}_{2-}$ are defined as follows, forming a set of complete bases of fermionic fields together with $\bar{\rho}_-^1, \bar{\tilde{\rho}}_{1+}$
\bal
\label{rotatedrho}
\Tilde{\rho}_-^1=-(\eta \bar{v})_a \Tilde{\psi}_-^a,\quad \Tilde{\rho}_+^2= (\bar{\eta} v)_a \Tilde{\psi}_+^a,&\quad \bar{\Tilde{\rho}}_{1+}=\epsilon^{ab} (\bar{\eta}v)_a \bar{\Tilde{\psi}}_{b+},\quad \bar{\Tilde{\rho}}_{2-}=\epsilon^{ab} (\eta \bar{v})_a \bar{\Tilde{\psi}}_{b-}\, .
\eal

Similar to the parametrization in 1/2 BPS Wilson loops, the constants $\bar{\delta}^{\dot{1}}$ and $\delta_{\dot{1}}$ (which are again not complex conjugate) satisfy $\delta_{\dot{1}} \bar{\delta}^{\dot{1}} =\frac{2i}{k}$. They are also allowed to depend on $\varphi$ with proper transformation of the shift, whose origin $\bar{\Gamma}$ in the superconnection is obtained in \cite{Drukker:2022ywj} to satisfy the relaxed supersymmtry condition \eqref{susycon}.
Depending on the factorisation of $\eta_a^{\imath}, \bar{\eta}_a^{\imath}$, the supersymmetries of the hyperloops in \eqref{Lfer} could vary from 1/4, 3/8 to 1/2 BPS. Analogous to the other 1/2 BPS loops \eqref{half2}, we can also construct the other hyperloops with the same gauge fields that preserve exactly the same symmetries as \eqref{Lfer}, with the replacements $\tilde{\rho}_+^{\dot{2}} \rightarrow \tilde{\rho}_-^{\dot{1}}$, $\bar{\tilde{\rho}}_{2-} \rightarrow -\bar{\tilde{\rho}}_{1+}$, $\bar{\Gamma}\rightarrow \bar{\Gamma} +\lambda/\Pi$ and the opposite sign for the $\tilde{\nu}$'s, compared to the ones appearing in \eqref{cAfer}.

In the following sections \ref{1/4 BPS Deformation} and \ref{3/8 BPS loops}, we study 1/4 and 3/8\footnote{We skip discussions about the 1/2 BPS case, because in principle it should be the same as the two-node hyperloops in \cite{Drukker:2022ywj} with exchange of hypermultiplets and twisted ones.} BPS \eqref{Lfer} respectively and their supersymmetric deformations.

\section{1/4 BPS hyperloops}
\label{1/4 BPS Deformation}
We notice that with the special choice of parameters in \eqref{lattitudeparameter}, the hyperloops in \eqref{Lfer} become the 1/4 BPS ``fermionic latitude" loops \cite{Asano:2012gt, Bianchi:2018bke, Drukker:2020dvr}, the supercharges preserved by which are not the subset of any 1/2 BPS loop. In other words, they are not in the moduli spaces of hyperloops obtained in \cite{Drukker:2022ywj}, so we can take them as the new starting point of supersymmetric deformation to discover new moduli spaces of hyperloops. 

However, since the latitude loop is just a special case of the general 1/4 BPS hyperloops in \eqref{Lfer} given by unfactorized $\eta_a^{\imath}$ and $\bar{\eta}_a^{\imath}$, we prefer to take the later ones as the starting point, in which case the preserved four supercharges are
\bal
\label{1/4supercharges}
\eta_a^{\imath} Q_{\imath}^{\dot{1}a+},\quad \eta_a^{\imath} Q_{\imath}^{\dot{2}a+},\quad \bar{\eta}_a^{\imath} (\sigma^1)_{\imath}^{\bar{\imath}} Q_{\bar{\imath}}^{\dot{1}a-},\quad \bar{\eta}_a^{\imath} (\sigma^1)_{\imath}^{\bar{\imath}} Q_{\bar{\imath}}^{\dot{2}a-}\, .
\eal
These supercharges make up the bases of a more general $\cQ$ with constant coefficients $w_{\dot{b}}$ and $\bar{w}_{\dot{b}}$ that carry $SU(2)_R$ indices
\bal
\label{Q1/4}
\cQ_{1/4}= w_{\dot{b}} \eta_a^{\imath} Q_{\imath}^{\dot{b}a+} +\bar{w}_{\dot{b}} \bar{\eta}_a^{\imath} (\sigma^1)_{\imath}^{\bar{\imath}} Q_{\bar{\imath}}^{\dot{b}a-}\, ,
\eal
where the subscript $1/4$ is written explicitly here to distinguish with the $3/8$ BPS case in the next section.

We now proceed to evaluate the action of this supercharge $\cQ_{1/4}$ on the superconnection $\cL_{1/4}$ in \eqref{Lfer} and to look for $H_{1/4}$ given in the total derivative \eqref{susycon}. To do so we define the rotated twisted scalar ﬁelds
\bal
\label{rotatedchi}
\Tilde{r}_{\dot{1}}=-\epsilon^{\dot{a}\dot{b}} w_{\dot{a}} \Tilde{q}_{\dot{b}},\quad \Tilde{r}_{\dot{2}}=\epsilon^{\dot{a}\dot{b}} \bar{w}_{\dot{a}} \Tilde{q}_{\dot{b}}, \quad \bar{\Tilde{r}}^{\dot{1}}=\bar{w}_{\dot{a}} \bar{\Tilde{q}}^{\dot{a}} ,\quad \bar{\Tilde{r}}^{\dot{2}}=w_{\dot{a}} \bar{\Tilde{q}}^{\dot{a}}\,.
\eal

It is useful to introduce a new parameter analogous to $\Pi$ \eqref{bos}
\bal
\label{tildePi}
\tilde{\Pi}= \epsilon^{\dot{a}\dot{b}} \bar{w}_{\dot{a}} w_{\bar{b}}\, ,
\eal
such that the supersymmetry variations of the rotated twisted scalar ﬁelds can be written in terms of it as
\bal
\label{Qchi}
\cQ_{1/4} \Tilde{r}_{\dot{1}} =\tilde{\Pi} \Tilde{\rho}_+^2,\quad \cQ_{1/4} \Tilde{r}_{\dot{2}} =\tilde{\Pi} \Tilde{\rho}_-^1,\quad \cQ_{1/4} \bar{\Tilde{r}}^{\dot{1}} =\tilde{\Pi} \bar{\Tilde{\rho}}_{2-},\quad \cQ_{1/4} \bar{\Tilde{r}}^{\dot{2}} =\tilde{\Pi} \bar{\Tilde{\rho}}_{1+}\, .
\eal
So far it is not hard to see the similarities between $\tilde{r}$ and $r$, $\tilde{\Pi}$ and $\Pi$. Moreover, using
\bal
\label{nu1/4}
\Tilde{r}_{I+1,\dot{1}} \bar{\Tilde{r}}_{I+1}^{\dot{1}} +\Tilde{r}_{I+1,\dot{2}} \bar{\Tilde{r}}_{I+1}^{\dot{2}}=\tilde{\Pi} \Tilde{\nu}_{I+1},\quad \bar{\Tilde{r}}_{I+1}^{\dot{1}} \Tilde{r}_{I+1,\dot{1}}+\bar{\Tilde{r}}_{I+1}^{\dot{2}} \Tilde{r}_{I+1,\dot{2}}=\tilde{\Pi} \Tilde{\nu}_{I+2}\,,
\eal
we get the second variations
\bal
\label{doubleQ1/4}
\cQ^2_{1/4} \Tilde{r}_{\dot{1}} &=\tilde{\Pi} \left(\Pi \left( i\partial_{\varphi} \Tilde{r}_{\dot{1}} +\Tilde{\cA}_{I+1} \Tilde{r}_{\dot{1}}- \Tilde{r}_{\dot{1}} \Tilde{\cA}_{I+2}\right) -\frac{1}{2} \epsilon^{ab} (\bar{\eta} v)_a (\eta \sigma^3 \bar{v})_b \Tilde{r}_{\dot{1}}\right)\\
\cQ^2_{1/4} \Tilde{r}_{\dot{2}} &=\tilde{\Pi} \left(\Pi \left( i\partial_{\varphi} \Tilde{r}_{\dot{2}} +\Tilde{\cA}_{I+1} \Tilde{r}_{\dot{2}}- \Tilde{r}_{\dot{2}} \Tilde{\cA}_{I+2}\right) -\frac{2i}{k}\Pi (\Tilde{\nu}_{I+1} \Tilde{r}_{\dot{2}} -\Tilde{r}_{\dot{2}} \Tilde{\nu}_{I+2}) -\frac{1}{2} \epsilon^{ab} (\bar{\eta} \sigma^3 v)_a (\eta \bar{v})_b \Tilde{r}_{\dot{2}}\right)\\
\cQ^2_{1/4} \bar{\Tilde{r}}^{\dot{1}} &=\tilde{\Pi} \left(\Pi \left( i \partial_{\varphi} \bar{\Tilde{r}}^{\dot{1}} + \Tilde{\cA}_{I+2} \bar{\Tilde{r}}^{\dot{1}} -\bar{\Tilde{r}}^{\dot{1}} \Tilde{\cA}_{I+1}\right) - \frac{1}{2} \epsilon^{ab} (\bar{\eta} \sigma^3 v)_a (\eta \bar{v})_b \bar{\Tilde{r}}^{\dot{1}}\right)\\
\cQ^2_{1/4} \bar{\Tilde{r}}^{\dot{2}} &=\tilde{\Pi} \left(\Pi \left( i \partial_{\varphi} \bar{\Tilde{r}}^{\dot{2}} + \Tilde{\cA}_{I+2} \bar{\Tilde{r}}^{\dot{2}} -\bar{\Tilde{r}}^{\dot{2}} \Tilde{\cA}_{I+1}\right) -\frac{2i}{k} \Pi (\Tilde{\nu}_{I+2} \bar{\Tilde{r}}^{\dot{2}} -\bar{\Tilde{r}}^{\dot{2}} \Tilde{\nu}_{I+1}) -\frac{1}{2} \epsilon^{ab} (\bar{\eta} v)_a (\eta \sigma^3 \bar{v})_b \bar{\Tilde{r}}^{\dot{2}}\right).
\eal

At this point all the supersymmetry variations of rotated twisted scalar fields have been figured out, we can then proceed to study the supersymmetric deformations of 1/4 BPS hyperloops in \eqref{Lfer}.

\subsection{Deformations with $\tilde{\Pi}\ne 0$}

We choose the starting point to be $\cL_{1/4}$ in \eqref{Lfer} preserving the supercharge $\cQ_{1/4}$ deﬁned in \eqref{Q1/4}. Following \cite{Drukker:2020dvr,Drukker:2022ywj}, we take a deformation of the form
\bal
\label{def1}
\cL=\cL_{1/4}-i\cQ_{1/4} G+B+C\,,
\eal
where $G$ is off-diagonal and Grassmann-even, so linear in the (twisted) scalar fields, $B$ is a diagonal bilinear and $C$ is annihilated by $\cQ_{1/4}$. BPS non-conformal loops with higher dimension insertions are also possible here, but again are not considered. On this condition, $\cQ_{1/4} C = 0$ splits into two cases: when the supercharge annihilates some of the matter ﬁelds and when it does not. As explained in \cite{Drukker:2022ywj}, we exclude the solutions in $C$ that includes any BPS bosonic loop where the supersymmetry variation is simply zero. Thus $C$ includes only matter fields from the twisted hypermultiplet that are annihilated by $\cQ_{1/4}$.

However, when $\tilde{\Pi}\ne0$, the only solutions to $\cQ_{1/4} C = 0$ which is at most bilinear in the ﬁelds is a numerical matrix containing no ﬁelds. So here we fix the gauge of deformation by setting $C=0$.

Moving to the off-diagonal $-i \cQ_{1/4} G$ term, we take
\bal
\label{G1/4}
G=\begin{pmatrix}
0 & \bar{b}^{\dot{a}} \Tilde{r}_{\dot{\dot{a}}}\\
b_{\dot{a}} \bar{\Tilde{r}}^{\dot{a}} & 0
\end{pmatrix}\,,
\eal
where the parameters $\bar{b}^{\dot{a}}, b_{\dot{a}}$ may be functions of $\varphi$.

By splitting the connection $\cL_{1/4}$ into the diagonal (bosonic) part $\cL_{1/4}^B$ and oﬀ-diagonal (fermionic) part $\cL_{1/4}^F$, the second supersymmetry variation of $G$ can be written as
\bal
\label{Q2G1/4}
-i \cQ_{1/4}^2 G =\partial_{\varphi} (\Pi \tilde{\Pi} G) -i[\cL_{1/4}^B, \Pi \tilde{\Pi} G]+i [H_{1/4}^2,G] -\hat{G}\,,
\eal
where we use $\cQ_{1/4} H_{1/4}=i \Pi \tilde{\Pi} \cL_{1/4}^F$ with $H_{1/4}$ given by $Q_{1/4} \cL =\cD_{\varphi}^{\cL} H_{1/4}$ in \eqref{susycon}
\bal
\label{HXine0}
H_{1/4}=\begin{pmatrix}
0 & \Pi \bar{\delta}^{\dot{1}} \Tilde{r}_{I+1,\dot{1}}\\
\delta_{\dot{1}} \bar{\Tilde{r}}_{I+1}^{\dot{1}} & 0
\end{pmatrix}\,,
\eal
and the remainder $\hat{G}$ is
\bal
\hat{G}=\begin{pmatrix}
0 & \Pi \tilde{\Pi} (\partial_{\varphi} \bar{b}^{\dot{a}}) \Tilde{r}_{\dot{a}} -i\lambda \tilde{\Pi} \bar{b}^{\dot{2}} \Tilde{r}_{\dot{2}}\\
\partial_{\varphi} (\Pi \tilde{\Pi} b_{\dot{a}}) \bar{\Tilde{r}}^{\dot{a}}+i\lambda \tilde{\Pi} b_{\dot{2}} \bar{\Tilde{r}}^{\dot{2}} & 0
\end{pmatrix}\,.
\eal

Since we are looking for the supersymmetric hyperloops with superconnections with the action of $\cQ_{1/4}$ on it to be a total derivative
\bal
\cQ_{1/4} \cL= \cD_{\varphi}^{\cL_{1/4}} H_{1/4} -i \cQ_{1/4}^2 G+ \cQ_{1/4} B=\cD_{\varphi}^{\cL} (H_{1/4} +\Delta H)\,,
\eal
by plugging in equation \eqref{Q2G1/4}, we get that $\Delta H=\Pi \tilde{\Pi} G$ and $B=\{H_{1/4},G\} + \Pi \tilde{\Pi} G^2$ on the condition $\hat{G}=0$. Consequently, almost the same as \cite{Drukker:2022ywj}, the deformation in form of \eqref{def1} is
\bal
\label{deformation}
\cL=\cL_{1/4}-i\cQ_{1/4} G+\{H_{1/4},G\} + \Pi \tilde{\Pi} G^2 +C\,.
\eal

The vanishing $\hat{G}$ gives four diﬀerential equations for $\bar{b}^{\dot{a}}$ and ${b}_{\dot{a}}$
\bal
i\partial_{\varphi} (\Pi {b}_{\dot{1}}) =0,&\qquad i\partial_{\varphi} (\Pi \bar{b}^{\dot{1}}) = (i\partial_{\varphi} \Pi) \bar{b}^{\dot{1}}\\
i\partial_{\varphi} (\Pi {b}_{\dot{2}}) =\lambda b_{\dot{2}},&\qquad i\partial_{\varphi} (\Pi \bar{b}^{\dot{2}}) = (i \partial_{\varphi} \Pi -\lambda) \bar{b}^{\dot{2}}\,.
\eal
As discussed in section 5.1 of \cite{Drukker:2022ywj}, for
\bal
\label{periodicalcondition}
\hat{c}(\varphi) =\int_0^{\varphi} \lambda/\Pi d\varphi^{\prime}\, ,
\eal
when $e^{i\hat{c}(\varphi)}$ is single valued, i.e. $e^{i\hat{c}(2\pi)} =1$, $\cL$ in \eqref{deformation} may couple to all twisted scalars, otherwise it could only couple either to the pair $\tilde{r}_{\dot{1}}, \bar{\tilde{r}}^{\dot{1}}$ or to $\tilde{r}_{\dot{2}}, \bar{\tilde{r}}^{\dot{2}}$. Note that with the unfactorised $\eta_a^{\imath}, \bar{\eta}_a^{\imath}$, $\lambda$ \eqref{lambda} is probably non-vanishing, so the periodical condition \eqref{periodicalcondition} is not guaranteed. However, we still focus on the case with two pairs of scalars, whose coefficients are given by
\bal
\bar{b}^{\dot{1}} =\frac{\bar{\beta}^{\dot{1}}-\bar{\delta}^{\dot{1}}}{\tilde{\Pi}},\quad \bar{b}^{\dot{2}} =\frac{\bar{\beta}^{\dot{2}} e^{i\hat{c}(\varphi)}}{\tilde{\Pi}},\quad b_{\dot{1}}=\frac{\beta_{\dot{1}}-\delta_{\dot{1}}}{\Pi \tilde{\Pi}},\quad b_{\dot{2}}=\frac{\beta_{\dot{2}} e^{-i\hat{c}(\varphi)}}{\Pi \tilde{\Pi}}\, ,
\eal
where $\bar{\beta}^{\dot{a}}, \beta_{\dot{a}}$ are constants, and to be consistent with the results in the later section \ref{3/8 BPS loops}, we put $\tilde{\Pi}$ in the denominator explicitly rather than absorb it into $\bar{\beta}^{\dot{a}}, \beta_{\dot{a}}$ since $\tilde{\Pi}$ is a constant parameter.

As a result, the deformed connection $\cL$ in \eqref{deformation} is
\bal
\label{defromedL1/41}
\begin{pmatrix}
\tilde{\cA}_{1/4,I+1}-\frac{1}{2} & -i \bar{\beta}^{\dot{1}} \Tilde{\rho}_{I+1,+}^2 -i \bar{\beta}^{\dot{2}} e^{i\hat{c}(\varphi)} \Tilde{\rho}_{I+1,-}^1\\
-i\frac{\beta_{\dot{1}}}{\Pi} \bar{\Tilde{\rho}}_{I+1,2-} -i\frac{\beta_{\dot{2}}}{\Pi} e^{-i\hat{c}(\varphi)} \bar{\Tilde{\rho}}_{I+1,1+} & \tilde{\cA}_{1/4,I+2} +\bar{\Gamma}
\end{pmatrix}\, ,
\eal
where
\bal
\tilde{\cA}_{1/4}=A_{\varphi} +\frac{i}{k\Pi} (r^1 \bar{r}_1 -r^2 \bar{r}_2) + \frac{1}{\tilde{\Pi}}\Tilde{M}_{\dot{b}}^{\dot{a}} \Tilde{r}_{\dot{a}} \bar{\Tilde{r}}^{\dot{b}}\, .
\eal
Note that the subscript $1/4$ means this is the deformation away from 1/4 BPS loops rather than $\tilde{\cA}_{1/4}$ itself preserves 1/4 BPS supersymmetry, and
\bal
\Tilde{M}_{\dot{b}}^{\dot{a}}=\begin{pmatrix}
-\frac{i}{k} +\bar{\beta}^{\dot{1}}\beta_{\dot{1}} & \bar{\beta}^{\dot{1}} \beta_{\dot{2}} e^{-i\hat{c}(\varphi)}\\
\bar{\beta}^{\dot{2}} \beta_{\dot{1}} e^{i\hat{c}(\varphi)} & \frac{i}{k} +\bar{\beta}^{\dot{2}}\beta_{\dot{2}}
\end{pmatrix}\, .
\eal
After ﬁxing a supercharge $\cQ_{1/4}$, the possible space of hyperloops is given by four complex parameters $\bar{\beta}^{\dot{a}}, \beta_{\dot{a}}$ modded by $\mathbb{C}^{\star}$, which is a conifold. This is the same type of moduli space found in \cite{Drukker:2019bev, Drukker:2020dvr,Drukker:2022ywj}, but totally independent from them except for some joint points, for example \eqref{Lfer}, the starting point of deformation. Generally such loops preserve only one supercharge \eqref{Q1/4} thus being 1/16 BPS, but there are some special cases where they receive certain supersymmetry enhancements as the discussion in the following.
\begin{itemize}
\item Single node bosonic loops.
    
We may decouple the nodes by simply setting $\bar{\beta}^{\dot{1}}=\bar{\beta}^{\dot{2}}=\beta_{\dot{1}}=\beta_{\dot{2}}=0$. This eliminates all the fermions in the superconnection \eqref{defromedL1/41}, thus it becomes block-diagonal with entries
\bal
\label{rotated1/8bos}
\cA^{\mathrm{bos}}_{1/8}=A_{\varphi} +\frac{i}{k \Pi} (r^1 \bar{r}_1 - r^2 \bar{r}_2) -\frac{i}{k \tilde{\Pi}} (\bar{\tilde{r}}^{\dot{1}} \tilde{r}_{\dot{1}} -\bar{\tilde{r}}^{\dot{2}} \tilde{r}_{\dot{2}})\,.
\eal
Under the condition that $\eta_a^{\imath},\bar{\eta}_a^{\imath}$ are unfactorised, these can only be 1/8 BPS bosonic loops, in a rotated version of \eqref{bos}. The two preserved loops are naturally the barred and unbarred parts of \eqref{Q1/4}. Noticing that now we can write the double supersymmetric variations of the twisted scalar fields in \eqref{doubleQ1/4} in terms of the bosonic loops
\bal
\cQ_{1/4}^2 \Tilde{r}_{\dot{1}} &=\tilde{\Pi} \left(\Pi \left( i\partial_{\varphi} \Tilde{r}_{\dot{1}} +\cA^{\mathrm{bos}}_{1/8,I+1} \Tilde{r}_{\dot{1}}- \Tilde{r}_{\dot{1}} \cA^{\mathrm{bos}}_{1/8,I+2}\right) -\frac{1}{2} \epsilon^{ab} (\bar{\eta} v)_a (\eta \sigma^3 \bar{v})_b \Tilde{r}_{\dot{1}}\right)\\
\cQ_{1/4}^2 \bar{\Tilde{r}}^{\dot{1}} &=\tilde{\Pi} \left(\Pi \left( i \partial_{\varphi} \bar{\Tilde{r}}^{\dot{1}} + \cA^{\mathrm{bos}}_{1/8,I+2} \bar{\Tilde{r}}^{\dot{1}} -\bar{\Tilde{r}}^{\dot{1}} \cA^{\mathrm{bos}}_{1/8,I+1}\right) - \frac{1}{2} \epsilon^{ab} (\bar{\eta} \sigma^3 v)_a (\eta \bar{v})_b \bar{\Tilde{r}}^{\dot{1}}\right)\,,
\eal
likewise for $\tilde{r}_{\dot{2}}, \bar{\Tilde{r}}^{\dot{2}}$. This implies us another path to obtain all the loops in the form of \eqref{defromedL1/41}, which is to deform from the bosonic loops \eqref{rotated1/8bos} with proper supercharge \eqref{Q1/4}. Actually, if we take
\bal
G=\begin{pmatrix}
    0 & \frac{1}{\tilde{\Pi}}(\bar{\beta}^{\dot{1}} \tilde{r}_{\dot{1}}+ \bar{\beta}^{\dot{2}} e^{i\hat{c}(\varphi)} \tilde{r}_{\dot{2}})\\
    \frac{1}{\Pi \tilde{\Pi}} (\beta_{\dot{1}} \bar{\tilde{r}}^{\dot{1}} +\beta_{\dot{2}} e^{-i\hat{c}(\varphi)} \bar{\tilde{r}}^{\dot{2}} &0
\end{pmatrix},\quad c=\bar{\Gamma} +\frac{1}{2}\, ,
\eal
with the deformation
\bal
\label{bosdef}
\diag(\cA_I^{\mathrm{bos}},\cA_{I+1}^{\mathrm{bos}}) \rightarrow \diag(\cA_I^{\mathrm{bos}},\cA_{I+1}^{\mathrm{bos}}+c)-i \cQ G+ \Pi \tilde{\Pi} G^2\, ,
\eal
we are able to recover arbitrary hyperloops in \eqref{defromedL1/41}.

\item Fermioic loops with non-trivial diagonal $\tilde{M}$.

When turning on only one pair of $\bar{\beta}, \beta$'s, for example $\bar{\beta}^{\dot{1}}$ and $\beta_{\dot{1}}$, we get diagonal $\tilde{M}$ and the resulting loops receive a natural supersymmetry enhancement. In addition to the supercharge in \eqref{Q1/4}, these loops
preserve at least the other supercharge 
\bal
\cQ_{1/4}^{\prime}=w_{\dot{b}} \eta_a^{\imath} Q_{\imath}^{\dot{b}a+} -\bar{w}_{\dot{b}} \bar{\eta}_a^{\imath} (\sigma^1)_{\imath}^{\bar{\imath}} Q_{\bar{\imath}}^{\dot{b}a-}\,.
\eal
To see this, in the case of $\cQ_{1/4}^{\prime}$ we change $\tilde{\Pi} \rightarrow -\tilde{\Pi}$ and $\tilde{M} \rightarrow -\tilde{M}$, with the same $\bar{\beta},\beta$'s the connection remains invariant. In particular, when $\bar{\beta}^{\dot{1}}\beta_{\dot{1}}=\frac{2i}{k}$ (or $\bar{\beta}^{\dot{2}}\beta_{\dot{2}}=-\frac{2i}{k}$ in the case $\bar{\beta}^{\dot{2}},\beta_{\dot{2}}$ are turned on), $\tilde{M}$ is proportional to the identity and restore $SU(2)_R$ R-symmetry, so the resulting loops become 1/4 BPS. 

\item ``Fermionic latitude” loops and ``Jordan normal form" fermionic loops.

Another special case is the ``fermionic latitude” loops constructed in \cite{Drukker:2020dvr}, which here can be recovered with parameters in \eqref{lattitudeparameter}. Besides the latitude loops and their conjugates up to $SU(2)_L\times SU(2)_R\times SL_2(\mathbb{R})$ which are the symmetries preserved by a great circle, similar analysis to \cite{Drukker:2022bff} tells that there is the other class of hyperloops, with $\eta_a^{\imath}$ or $\bar{\eta}_a^{\imath}$ in the non-trivial Jordan normal forms \eqref{Jordannormal}. With such choices of preserved supercharges, \eqref{defromedL1/41} can be viewed as fermionic deformations of \eqref{JNbosloops}.
\end{itemize}

\subsection{Deformations with $\Tilde{\Pi}=0$}
\label{Xi=0}
The other possible case is for
\bal
\tilde{\Pi}= \bar{w}_{\dot{1}} w_{\dot{2}} - \bar{w}_{\dot{2}}w_{\dot{1}} =0\, ,
\eal
where $\cQ_{1/4}$ has a nontrivial kernel \eqref{rotatedchi} (see \eqref{Qchi} for the short proof), that brings out some novel cases. Although $\tilde{\Pi}=0$, the four parameters $w_{\dot{a}},\bar{w}_{\dot{a}}$ can not be all zero, otherwise the supercharge \eqref{Q1/4} just vanishes. We assume $\bar{w}_{\dot{1}}\ne 0$, and introduce the quotient parameter
\bal
\label{omega}
\omega=\frac{w_{\dot{1}}}{\bar{w}_{\dot{1}}}=\frac{w_{\dot{2}}}{\bar{w}_{\dot{2}}}\,,
\eal
where the constant $\omega \in \mathbb{C}\cup \{\infty\}$\footnote{$\omega=\infty$ corresponds to $\bar{w}_{\dot{1}}= \bar{w}_{\dot{2}} =0$, which is the case we avoid discussing where only the unbarred supercharges are preserved. Similarly for $\omega=0$.}.

The pairs of rotated ﬁelds defined in \eqref{rotatedchi} are not linearly independent
\bal
\Tilde{r}_{\dot{1}}=-\omega \Tilde{r}_{\dot{2}},\quad \bar{\Tilde{r}}^{\dot{2}} =\omega \bar{\Tilde{r}}^{\dot{1}}\,.
\eal
So in order to construct a new basis of the twisted scalar ﬁelds, besides
\bal
\Tilde{r}_{\parallel} =-\Tilde{r}_{\dot{2}},\quad \bar{\Tilde{r}}^{\parallel}=\bar{\Tilde{r}}^{\dot{1}}\,,
\eal
we also need an orthogonal pair which are not annihilated by $\cQ$. Define
\bal
\Tilde{r}_{\perp} =\bar{w}_{\dot{1}} \Tilde{q}_{\dot{1}} +\bar{w}_{\dot{2}} \Tilde{q}_{\dot{2}},\quad \bar{\Tilde{r}}^{\perp} =\bar{w}_{\dot{2}} \bar{\Tilde{q}}^{\dot{1}} -\bar{w}_{\dot{1}} \bar{\Tilde{q}}^{\dot{2}}\,,
\eal
so that similar to \eqref{nu1/4}, there are
\bal
\label{nu1/40}
\Tilde{r}_{I+1,\parallel} \bar{\Tilde{r}}_{I+1}^{\perp} + \Tilde{r}_{I+1,\perp} \bar{\Tilde{r}}_{I+1}^{\parallel} =\bar{\tilde{\Lambda}} \Tilde{\nu}_{I+1},\quad \bar{\Tilde{r}}_{I+1}^{\perp} \Tilde{r}_{I+1,\parallel} + \bar{\Tilde{r}}_{I+1}^{\parallel} \Tilde{r}_{I+1,\perp}=\bar{\tilde{\Lambda}} \Tilde{\nu}_{I+2}\,,
\eal
where
\bal
\bar{\tilde{\Lambda}}=\bar{w}_{\dot{1}}^2 +\bar{w}_{\dot{2}}^2\,.
\eal
Since  $w_{\dot{a}},\bar{w}_{\dot{a}}$ can not be all zero, $\bar{\tilde{\Lambda}}$ is always nonvanishing. With it we get single and double supersymmetric transformations of $\Tilde{\chi}_{\perp}, \bar{\tilde{\chi}}^{\perp}$
\bal
\cQ_{1/4} \Tilde{r}_{\perp} &=\bar{\tilde{\Lambda}} (\omega \Tilde{\rho}_-^1+ \Tilde{\rho}_+^2),\quad \cQ_{1/4} \bar{\Tilde{r}}^{\perp}=\bar{\tilde{\Lambda}} ( \omega \bar{\Tilde{\rho}}_{2-} -\bar{\Tilde{\rho}}_{1+})\\
\cQ_{1/4}^2 \Tilde{r}_{\perp} &=\bar{\tilde{\Lambda}} \left( \omega \lambda \Tilde{\chi}_{\parallel} +\frac{2i}{k}\omega \Pi (\nu_{I+1} \Tilde{\chi}_{\parallel} - \Tilde{\chi}_{\parallel} \nu_{I+2}) \right)\\
\cQ_{1/4}^2 \bar{\Tilde{r}}^{\perp} &= -\bar{\tilde{\Lambda}} \left( \omega \lambda \bar{\Tilde{\chi}}^{\parallel} -\frac{2i}{k}\omega \Pi (\nu_{I+2} \bar{\Tilde{\chi}}^{\parallel} -\bar{\Tilde{\chi}}^{\parallel} \nu_{I+1}) \right)\,.
\eal

Now we proceed to reconsider the deformation \eqref{def1}, for which the same formalism \eqref{deformation} as in the $\tilde{\Pi}\ne0$ case can be applied
\bal
\label{deformationXi=0}
\cL=\cL_{1/4}-i\cQ_{1/4} G+\{H_{1/4},G\} +C,\quad \cQ_{1/4} C=0\,,
\eal
where $H$ is the same as above \eqref{HXine0}, just in the new notations it becomes
\bal
H_{1/4}=\begin{pmatrix}
0 & \omega \Pi \bar{\delta}_{I+1}^{\dot{1}} \Tilde{r}_{I+1,\parallel}\\
\delta_{I+1,\dot{1}} \bar{\Tilde{r}}_{I+1}^{\parallel} & 0
\end{pmatrix}\,.
\eal
The action of $\cQ_{1/4}$ on the superconnection $\cL$ will again be a total derivative
\bal
\cQ_{1/4} \cL=\cD_{\varphi}^{\cL} H_{1/4}\,.
\eal

The most significant distinction between \eqref{deformationXi=0} and the previous case is $C$. Because of the fact that the supercharge $\cQ_{1/4}$ in \eqref{Q1/4} annihilates the rotated twisted scalars $\Tilde{\chi}_{\parallel}, \bar{\Tilde{\chi}}^{\parallel}$, $C$ may contain their bilinears as well as the numerical factor $c$, explicitly
\bal
\label{C1/4}
C=\begin{pmatrix}
\bar{\beta}^{\parallel} \Tilde{r}_{\parallel} \bar{\Tilde{r}}^{\parallel} & 0 \\
0 & \beta_{\parallel} \bar{\Tilde{r}}^{\parallel} r_{\parallel} +c
\end{pmatrix}\,.
\eal

The new off-diagonal matrix $G$ is comprised of the other pair of the rotated twisted scalar fields only
\bal
\label{G01/4}
G=\begin{pmatrix}
0 & \bar{\beta}^{\perp} \Tilde{r}_{\perp}\\
\beta_{\perp} \bar{\Tilde{r}}^{\perp} & 0
\end{pmatrix}\,,
\eal
where the parameters $\bar{\beta}^{\perp},\beta_{\perp}$ may be functions of $\varphi$. If the parmeters satisfy
\bal
\label{Xi0eqn}
\bar{\tilde{\Lambda}} \omega \lambda \bar{\beta}^{\perp}=-c \omega \Pi \bar{\delta}^{\dot{1}},\quad \bar{\tilde{\Lambda}} \omega \lambda \beta_{\perp} =-c \delta_{\dot{1}},\quad \bar{\beta}^{\parallel} =\beta_{\parallel}\,,
\eal
The resulting supersymmetric loops obtained by \eqref{deformationXi=0} are
\bal
\label{Pi=0loops1/4}
\begin{pmatrix}
\tilde{\cA}_{1/4,I+1}^0 -\frac{1}{2} & -i\bar{\tilde{\Lambda}} \omega \bar{\beta}^{\perp} \Tilde{\rho}_{I+1,-}^1 -i (\bar{\delta}^{\dot{1}} +\bar{\tilde{\Lambda}} \bar{\beta}^{\perp})\Tilde{\rho}_{I+1,+}^2\\
i \bar{\tilde{\Lambda}} \beta_{\perp} \bar{\Tilde{\rho}}_{I+1,1+} -i(\frac{\delta_{\dot{1}}}{\Pi}+ \bar{\tilde{\Lambda}} \omega \beta_{\perp}) \bar{\Tilde{\rho}}_{I+1,2-} & \tilde{\cA}_{1/4,I+2}^0 +\bar{\Gamma} +c
\end{pmatrix}\, ,
\eal
where
\bal
\tilde{\cA}_{1/4}^0=A_{\varphi} +\frac{i}{k\Pi} (r^1 \bar{r}_{1} -r^2 \bar{r}_2) +\Tilde{M}_{\dot{b}}^{\dot{a}} \Tilde{r}_{\dot{a}} \bar{\Tilde{r}}^{\dot{b}}\,,
\eal
with $\dot{a},\dot{b}=\perp,\parallel$ and
\bal
\Tilde{M}_{\dot{b}}^{\dot{a}}=\begin{pmatrix}
0 & \frac{i}{k \bar{\tilde{\Lambda}}} +\bar{\beta}^{\perp} \delta_{\dot{1}}\\
\frac{i}{k \bar{\tilde{\Lambda}}} +\omega \Pi \bar{\delta}^{\dot{1}} \beta_{\perp} & \bar{\beta}^{\parallel}
\end{pmatrix}\,,
\eal
with $\beta_{\perp},\bar{\beta}^{\perp},c$ solutions of \eqref{Xi0eqn}. Because of the unfactorised $\eta_a^{\imath}, \bar{\eta}_a^{\imath}$, the resulting loops are independent from the $\Pi=0$ ones in \cite{Drukker:2022ywj}, in which case the factorisation of $\eta_a^{\imath}, \bar{\eta}_a^{\imath}$ is required automatically by the condition $\Pi=0$. Generically, these loops only preserve one supercharge, while just like $\Pi \ne 0$ case at some special points again we ﬁnd supersymmetry enhancements.

It is shown in \eqref{Xi0eqn} that the solutions of parameters $\beta,\bar{\beta}$ with $\perp$ and $\parallel$ indices are independent from each other. The simplest case for loops with enhanced supersymmetry is when the superconnection is invariant under $SU(2)_R$, i.e. when $\bar{\beta}^{\parallel}=0$ and
$\bar{\beta}^{\perp} \delta_{\dot{1}}= \omega \Pi \bar{\delta}^{\dot{1}} \beta_{\perp}$, which is consistent with the solutions of \eqref{Xi0eqn}. Since $\eta_a^{\imath}$ and $\bar{\eta}_a^{\imath}$ are unfactorized, it turns out this is the only enhancement that the resulting loops receive, thus being 1/8 BPS. The preserved supercharges are
\bal
\omega \eta_a^{\imath} Q_{\imath}^{\dot{1}a} +\bar{\eta}_a^{\imath} (\sigma^1)_{\imath}^{\bar{\imath}} Q_{\bar{\imath}}^{\dot{1}a},\quad \omega \eta_a^{\imath} Q_{\imath}^{\dot{2}a} +\bar{\eta}_a^{\imath} (\sigma^1)_{\imath}^{\bar{\imath}} Q_{\bar{\imath}}^{\dot{2}a}\, .
\eal

Two further degenerations are when $\omega$ vanishes or goes to inﬁnite. In both cases the preserved supercharges are composed entirely of either barred or unbarred ones which as mentioned in the introduction, are beyond the discussion of this paper.

\section{3/8 BPS hyperloops}
\label{3/8 BPS loops}
When one of $\eta_a^{\imath},\bar{\eta}_a^{\imath}$ factorises while the other does not, the hyperloops in \eqref{Lfer} are enhanced to 3/8 BPS. As an example, we focus on the case where the factorised ones are $\bar{\eta}_a^{\imath}$, i.e. $\bar{\eta}_a^{\imath} =\bar{y}^{\imath} \bar{u}_a$. Following the definition of $\Pi$ in \eqref{tildePi}, now it turns into $\Pi=(\bar{y} v) \epsilon^{ab} \bar{u}_a (\eta \bar{v})_b$, where the same factor $\bar{y} v$ appears also in $r^2$ and $\bar{r}_1$ \eqref{rotatedscalars}. This allows us to define the corresponding ``factorised" parameters and rotated fields without such a common factor
\bal
\label{rf}
\Pi^f=\epsilon^{ab} \bar{u}_a (\eta \bar{v})_b, \quad r^{f2}=\bar{u}_a q^a,\quad \bar{r}_1^f=\epsilon^{ab} \bar{u}_a \bar{q}_b\, ,
\eal
so that the bilinears of untwisted fields in \eqref{cAfer} remains unchanged with factorised $\Pi^f$
\bal
\frac{1}{\Pi^f} (r^1 \bar{r}_1^f -r^{f2} \bar{r}_2)=\frac{1}{\Pi} (r^1 \bar{r}_1 -r^2 \bar{r}_2)\,.
\eal
However, if we furthermore want to express the fermions in \eqref{Lfer} with factorised parameters as well, because of partial derivatives in the supersymmtric condition \eqref{susycon}, the shift part in the connection has to be 0 adaptively. Explicitly
\bal
\label{L3/8}
\cL_{3/8}=\begin{pmatrix}
    \tilde{\cA}_{\varphi,I+1} -\frac{1}{2} & -i \bar{\delta}^{\dot{1}} \tilde{\rho}_+^{2f} \\
    -i \frac{\delta_{\dot{1}} }{\Pi^f} \bar{\tilde{\rho}}_{2-} & \tilde{\cA}_{\varphi,I+2}
\end{pmatrix}\,,
\eal
where the factorised rotated fermions $\tilde{\rho}^{f2}_+$ is defined by $\tilde{\rho}^{f2}_+ =\bar{u}_a \tilde{\psi}_+^a$. Note that since \eqref{L3/8} is just a $U(1)$ symmetry transformed version of \eqref{Lfer}, we can also take the later one directly with factorised $\bar{\eta}_a^{\imath}$ and unfactorised $\eta_a^{\imath}$, as well as the original rotated fields defined in \eqref{rotatedscalars}, \eqref{rotatedrho} and \eqref{rotatedchi}. Besides, the periodical condition \eqref{periodicalcondition} is always satisfied, though the shift is generally non-zero. The real reason that drives us to turn to \eqref{L3/8} rather than sticking with \eqref{Lfer} is that instead of the four preserved supercharges in \eqref{1/4supercharges}, now we have six 
\bal
\label{6supercharges}
\eta_a^{\imath} Q_{\imath}^{\dot{1}a},\quad \eta_a^{\imath} Q_{\imath}^{\dot{2}a},\quad \bar{u}_a Q^{\dot{1}a}_{\bar{l}},\quad \bar{u}_a Q^{\dot{1}a}_{\bar{r}},\quad \bar{u}_a Q^{\dot{2}a}_{\bar{l}},\quad \bar{u}_a Q^{\dot{2}a}_{\bar{r}}\,,
\eal
which can be packaged into a superposition
\bal
\label{cQ3/8}
\cQ_{3/8}=w_{\dot{b}} \eta_a^{\imath} Q_{\imath}^{\dot{b}a} +\bar{w}_{\dot{b}}^{\imath} \bar{u}_a (\sigma^1)_{\imath}^{\bar{\imath}} Q^{\dot{b}a}_{\bar{\imath}}\,.
\eal
Instead, the factorised version of 1/4 BPS supercharge \eqref{Q1/4} is just $\cQ_{1/4}^f= w_{\dot{b}} \eta_a^{\imath} Q_{\imath}^{\dot{b}a+} +\bar{w}_{\dot{b}} \bar{y}^{\imath} \bar{u}_a (\sigma^1)_{\imath}^{\bar{\imath}} Q_{\bar{\imath}}^{\dot{b}a-}$, which is identical to \eqref{cQ3/8} if and only if $\bar{w}_{\dot{b}}^{\imath}$ factorises. So to be more general, we would like to consider both of the cases with factorised and unfactorised $\bar{w}_{\dot{b}}^{\imath}$, and the proper gauge we should choose is the one in which superconnection can be written as \eqref{L3/8}.

There are other fields and parameters that are ``factorised" as well along with the supercharges
\bal
\tilde{r}_{\dot{2}}^f=\epsilon^{\dot{a}\dot{b}} (\bar{w} v)_{\dot{a}}\tilde{q}_{\dot{b}},\quad \bar{\tilde{r}}^{f \dot{1}} =(\bar{w} v)_{\dot{a}} \bar{\tilde{q}}^{\dot{a}},\quad \bar{\tilde{\rho}}_{1+}^f= \epsilon^{ab} \bar{u}_a \bar{\tilde{\psi}}_{b+},\quad \tilde{\Pi}^f= \epsilon^{\dot{a} \dot{b}} (\bar{w}v)_{\dot{a}} w_{\dot{b}}\,.
\eal
Loosely summarizing, the phases $v,\bar{v}$ dropping out of $\bar{\eta}$ are picked up by $\bar{w}$ in this case. In this way, \eqref{nu1/4} now becomes
\bal
\label{nu3/8}
\Tilde{r}_{I+1,\dot{1}} \bar{\Tilde{r}}_{I+1}^{f\dot{1}} +\Tilde{r}_{I+1,\dot{2}}^f \bar{\Tilde{r}}_{I+1}^{\dot{2}}=\tilde{\Pi}^f \Tilde{\nu}_{I+1},\quad \bar{\Tilde{r}}_{I+1}^{f\dot{1}} \Tilde{r}_{I+1,\dot{1}}+\bar{\Tilde{r}}_{I+1}^{\dot{2}} \Tilde{r}_{I+1,\dot{2}}^f =\tilde{\Pi}^f \Tilde{\nu}_{I+2}\,.
\eal

Analogous to \eqref{Qchi}, in the factorised notation, the supersymmetry variations of the scalar fields are
\bal
\cQ_{3/8} \tilde{r}_{\dot{1}} =\tilde{\Pi}^f \tilde{\rho}_+^{f2},\quad \cQ_{3/8} \tilde{r}_{\dot{2}}^f =\tilde{\Pi}^f \tilde{\rho}_-^{1},\quad \cQ_{3/8} \bar{\tilde{r}}^{f\dot{1}} =\tilde{\Pi}^f \bar{\tilde{\rho}}_{2-},\quad \cQ_{3/8} \bar{\tilde{r}}^{\dot{2}} =\tilde{\Pi}^f \bar{\tilde{\rho}}_{1+}^f\,,
\eal
and the second variations
\bal
\cQ_{3/8}^2 \tilde{r}_{\dot{1}} &=\tilde{\Pi}^f \left( \Pi^f \left( i\partial_{\varphi} \tilde{r}_{\dot{1}} +\tilde{\cA}_{I+1} \tilde{r}_{\dot{1}} - \tilde{r}_{\dot{1}} \tilde{\cA}_{I+2} \right) -\frac{1}{2}\epsilon^{ab} \bar{u}_a (\eta \sigma^3 \bar{v})_b \tilde{r}_{\dot{1}} \right)\\
\cQ_{3/8}^2 \tilde{r}_{\dot{2}}^f &=\tilde{\Pi}^f \left( \Pi^f \left(i\partial_{\varphi} \tilde{r}_{\dot{2}}^f +\tilde{\cA}_{I+1} \tilde{r}_{\dot{2}}^f - \tilde{r}_{\dot{2}}^f \tilde{\cA}_{I+2} +\frac{1}{2} \tilde{r}_{\dot{2}}^f \right) -\frac{2i}{k} \Pi^f (\tilde{\nu}_{I+1} \tilde{r}_{\dot{2}}^f - \tilde{r}_{\dot{2}}^f \tilde{\nu}_{I+2}) \right)\\
\cQ_{3/8}^2 \bar{\tilde{r}}^{f\dot{1}} &=\tilde{\Pi}^f \left( \Pi^f \left(i\partial_{\varphi}  \bar{\tilde{r}}^{f\dot{1}} +\tilde{\cA}_{I+2} \bar{\tilde{r}}^{f\dot{1}} - \bar{\tilde{r}}^{f\dot{1}} \tilde{\cA}_{I+1} +\frac{1}{2} \bar{\tilde{r}}^{f\dot{1}}\right) \right)\\
\cQ_{3/8}^2 \bar{\tilde{r}}^{\dot{2}} &=\tilde{\Pi}^f \left( \Pi^f \left(i\partial_{\varphi} \bar{\tilde{r}}^{\dot{2}} +\tilde{\cA}_{I+2} \bar{\tilde{r}}^{\dot{2}} - \bar{\tilde{r}}^{\dot{2}} \tilde{\cA}_{I+1} \right)\! -\!\frac{2i}{k} \Pi^f (\tilde{\nu}_{I+2} \bar{\tilde{r}}^{\dot{2}} - \bar{\tilde{r}}^{\dot{2}} \tilde{\nu}_{I+1}) -\frac{1}{2} \epsilon^{ab} \bar{u}_a (\eta \sigma^3 \bar{v})_b \bar{\tilde{r}}^{\dot{2}} \!\right)\!.
\eal

Then we can easily check that similar to the 1/2 and 1/4 cases, here we have $\cQ_{3/8} H_{3/8} =i \Pi^f \tilde{\Pi}^f \cL_{3/8}^F$ with
\bal
\label{H3/8}
H_{3/8}=\begin{pmatrix}
    0 & \bar{\delta}^{\dot{1}} \Pi^f \tilde{r}_{\dot{1}} \\
    \delta_{\dot{1}} \bar{\tilde{r}}^{f\dot{1}} & 0
\end{pmatrix}
\eal
which is  equal to the factorised \eqref{HXine0}.

\subsection{Deformations with $\tilde{\Pi}^f \ne 0$}
A general supersymmetric deformation from the 3/8 BPS loops \eqref{L3/8} is totally the same as \eqref{deformation}, except that everything is in the factorised notation now
\bal
\label{deformation3/8}
\cL=\cL_{3/8} -i\cQ_{3/8} G+\{H_{3/8},G\} +\Pi^f \tilde{\Pi}^f G^2 +C\,,
\eal
where we take same coefficients $\bar{b},b$ in $G$ as in \eqref{G1/4}, but of the factorised rotated fields $\tilde{r}_{\dot{1}}, \tilde{r}_{\dot{2}}^f, \bar{\tilde{r}}^{f \dot{1}}$ and $\bar{\tilde{r}}^{\dot{2}}$. Again we have $\cQ_{3/8}$ acting on $\cL$ to be a covariant derivative
\bal
\cQ_{3/8} \cL= \cD_{\varphi}^{\cL} (H_{3/8} +\Pi^f \tilde{\Pi}^f G)\,,
\eal
as long as
\bal
i\partial_{\varphi} (\Pi^f \tilde{\Pi}^f b_{\dot{1}})=0,\qquad &i\partial_{\varphi} (\tilde{\Pi}^f \bar{b}^{\dot{1}})=0\\
i\partial_{\varphi} (\tilde{\Pi}^f b_{\dot{2}})=-\tilde{\Pi}^f b_{\dot{2}},\qquad& i\partial_{\varphi} (\Pi^f \tilde{\Pi}^f \bar{b}^{\dot{2}}) =\Pi^f \tilde{\Pi}^f \bar{b}^{\dot{2}}\,.
\eal

The solutions are
\bal
\bar{b}^{\dot{1}}= \frac{\bar{\beta}^{\dot{1}} -\bar{\delta}^{\dot{1}}}{\tilde{\Pi}^f},\quad \bar{b}^{\dot{2}}=\frac{\bar{\beta}^{\dot{2}}e^{-i\varphi}}{\Pi^f \tilde{\Pi}^f},\quad b_{\dot{1}} =\frac{\beta_{\dot{1}}-\delta_{\dot{1}}}{\Pi^f \tilde{\Pi}^f},\quad b_{\dot{2}} =\frac{\beta_{\dot{2}} e^{i\varphi}}{\tilde{\Pi}^f}\,,
\eal
which leads to supersymmetric loops that preserve \eqref{cQ3/8}
\bal
\label{defored3/8}
\begin{pmatrix}
    \tilde{\cA}_{3/8,I+1}-\frac{1}{2} & -i \bar{\beta}^{\dot{1}} \tilde{\rho}_+^{f2} -i \frac{\bar{\beta}^{\dot{2}} e^{-i\varphi}} {\Pi^f} \tilde{\rho}_-^1\\
    -i \frac{\beta_{\dot{1}}}{\Pi^f} -i\beta_{\dot{2}} e^{i\varphi} \bar{\tilde{\rho}}_{1+}^f & \tilde{\cA}_{3/8,I+2}
\end{pmatrix}\,,
\eal
where
\bal
\label{deformedL3/81}
\tilde{\cA}_{3/8} =A_{\varphi} +\frac{i}{k \Pi^f} (r^1 \bar{r}_1^f -r^{f2} \bar{r}_2) +\frac{1}{\tilde{\Pi}^f} \tilde{M}^{f\dot{a}}_{\dot{b}} \tilde{r}^{(f)}_{\dot{a}} \bar{\tilde{r}}^{(f)\dot{b}}\,,
\eal
with
\bal
\tilde{M}^{f\dot{a}}_{\dot{b}} =\begin{pmatrix}
    -\frac{i}{k} +\bar{\beta}^{\dot{1}} \beta_{\dot{1}} & \Pi^f \bar{\beta}^{\dot{1}} \beta_{\dot{2}} e^{i\varphi}\\
    \frac{1}{\Pi^f} \beta_{\dot{1}} \bar{\beta}^{\dot{2}} e^{-i\varphi} & \frac{i}{k} +\bar{\beta}^{\dot{2}} \beta_{\dot{2}}
\end{pmatrix}\,.
\eal

After fixing $\cQ_{3/16}$, the moduli space of hyperloops in \eqref{defored3/8} is another conifold independent from \eqref{defromedL1/41} and those in the previous references \cite{Drukker:2019bev, Drukker:2020dvr,Drukker:2022ywj}. They are somehow at an unexplored state between the deformed 1/4 BPS and 1/2 BPS loops.
The special points with supersymmetric enhancement are very similar to those deforming from 1/4 BPS loops \eqref{Lfer}, while allowing more possibilities. An example is the novel ``Factorised $w$ and $\bar{z}$" bosonic loops found in \cite{Drukker:2022bff} are included in the moduli space of \eqref{defored3/8}. Explicitly, from the choice $\bar{\beta}^{\dot{1}}=\bar{\beta}^{\dot{2}}=\beta_{\dot{1}}= \beta_{\dot{2}}=0$, we get single node bosonic loops
\bal
\label{bos3/16?}
A_{3/16?}^{\mathrm{bos}}= A_{\varphi} +\frac{i}{k \Pi^f} (r^1 \bar{r}_1^f -r^{f2} \bar{r}_2) -\frac{i}{k\tilde{\Pi}^f} (\tilde{r}_{\dot{1}} \bar{\tilde{r}}^{f\dot{1}} -\tilde{r}_{\dot{2}}^f \bar{\tilde{r}}^{\dot{2}})\,.
\eal
Compared to the 1/8 BPS bosonic loops in \eqref{rotated1/8bos}, we might think that the resulting loops are 3/16 BPS naively. However, this is only true when $\bar{w}_b^{\imath}$ factorise. If not, since $\eta_a^{\imath}$ is unfactorised, these loops are also 1/8 BPS, preserving the barred and unbarred supercharges in \eqref{cQ3/8} separately. Especially, taking $\bar{u}_a=\delta_a^1,\eta_1^l=\eta_2^r=1, w_a=\delta_a^2,\bar{w}_1^l=1,\bar{w}_2^r=-1$, we recover the ``Factorised $w$ and $\bar{z}$" 1/8 BPS bosonic loop.

Again by deforming the bosonic loops above, we are able to obtain any fermionic loops in \eqref{deformedL3/81}. This also implies us that the fermionic partners of the ``Factorised $w$ and $\bar{z}$" loops are included in our discussion.

\subsection{Deformations with $\tilde{\Pi}^f=0$}
\label{3/80}
The other case is for
\bal
\tilde{\Pi}^f =(\bar{w} v)_{\dot{1}} w_{\dot{2}} -(\bar{w} v)_{\dot{2}} w_{\dot{1}}=0\, .
\eal
Similar to \eqref{omega}, we define the quotient parameter
\bal
\label{omegaf}
\omega^f=\frac{(\bar{w}v)_{\dot{1}}}{w_{\dot{1}}} =\frac{(\bar{w}v)_{\dot{2}}}{w_{\dot{2}}}\, .
\eal
However, instead of a constant in \eqref{omega}, here it takes the form $a+b e^{i\varphi}$ with $a,b$ free constant parameters. Since $\bar{\tilde{r}}^{f\dot{1}},\tilde{r}_{\dot{1}}$ are linearly dependant with $\bar{\tilde{r}}^{\dot{2}}, \tilde{r}_{\dot{2}}^f$ in this case, we construct new orthogonal bases
\bal
\tilde{r}_{\parallel}^f =\tilde{r}_{\dot{1}},\quad \bar{\tilde{r}}^{f\parallel} =\bar{\tilde{r}}^{\dot{2}},\quad
\tilde{r}_{\perp}^f =w_{\dot{1}} \tilde{q}_{\dot{1}} +w_{\dot{2}} \tilde{q}_{\dot{2}},\quad \bar{\tilde{r}}^{f\perp} =w_{\dot{2}} \bar{\tilde{q}}^{\dot{1}} -w_{\dot{1}} \bar{\tilde{q}}^{\dot{2}}\,,
\eal
where the scalars with parallel indices are annihilated by the supercharge \eqref{cQ3/8}. Define
\bal
\tilde{\Lambda} =w_{\dot{1}}^2 +w_{\dot{2}}^2\,,
\eal
in terms of which similar identities as \eqref{nu1/40} exist, and supersymmetric variations of the new scalar bases are
\bal
\cQ_{3/8} \tilde{r}_{\perp}^f &=\tilde{\Lambda} (\tilde{\rho}_-^1 +\omega^f \tilde{\rho}_+^{f2}),\quad \cQ_{3/8} \bar{\tilde{r}}^{f\perp} =\tilde{\Lambda} (\bar{\tilde{\rho}}_{2-} -\omega^f \bar{\tilde{\rho}}_{1+}^f)\\
\cQ_{3/8}^2 \tilde{r}_{\perp}^f &=-\tilde{\Lambda} \Pi^f \omega^f \left( (i \partial_{\varphi} \log \frac{\omega^f}{\Pi^f} +1) \tilde{r}_{\parallel}^f -\frac{2i}{k} (\tilde{\nu}_{I+1} \tilde{r}_{\parallel}^f -\tilde{r}_{\parallel}^f \tilde{\nu}_{I+2}) \right)\\
\cQ_{3/8}^2 \bar{\tilde{r}}^{f\perp} &=\tilde{\Lambda} \omega^f \Pi^f \left( (i \partial_{\varphi} \log \frac{\omega^f}{\Pi^f} +1)\bar{\tilde{r}}^{f\parallel} +\frac{2i}{k} (\tilde{\nu}_{I+2} \bar{\tilde{r}}^{f\parallel} - \bar{\tilde{r}}^{f\parallel} \tilde{\nu}_{I+1})\right)\,.
\eal

The deformation is the same as \eqref{deformation3/8} with $\tilde{\Pi}^f=0$, where $H$ is also \eqref{H3/8} in the replacement of $\tilde{r}_{\dot{1}} \rightarrow \tilde{r}_{\parallel},\;\bar{\tilde{r}}^{f\dot{1}} \rightarrow \omega^f \bar{\tilde{r}}^{\parallel}$. With $\bar{\beta}^{\perp},\bar{\beta}_{\perp}$ satisfying
\bal
\label{sol3/8Pi=0}
c \bar{\delta}^{\dot{1}} \Pi^f =-i \bar{\beta}^{\perp} \tilde{\Lambda} \Pi^f \omega^f \partial_{\varphi} \log \left( \frac{\omega^f} {\Pi^f} e^{-i\varphi}\right),\quad c \delta_{\dot{1}} \omega^f =-i \beta_{\perp} \tilde{\Lambda} \Pi^f \omega^f \partial_{\varphi} \log \left( \frac{\omega^f} {\Pi^f} e^{-i\varphi}\right)\,,
\eal
the resulting loops are
\bal
\begin{pmatrix}
    \tilde{\cA}_{3/8,I+1}^0 -\frac{1}{2} & -i\tilde{\Lambda} \bar{\beta}^{\perp} \tilde{\rho}_-^1 -i(\bar{\delta}^{\dot{1}} +\tilde{\Lambda} \omega^f \bar{\beta}^{\perp}) \tilde{\rho}_+^{f2}\\
    i \tilde{\Lambda} \omega^f \beta_{\perp} \bar{\tilde{\rho}}_{1+}^f -i (\frac{\delta_{\dot{1}}}{\Pi^f} +\tilde{\Lambda} \beta_{\perp}) \bar{\tilde{\rho}}_{2-} & \tilde{\cA}_{3/8,I+2}^0+c
\end{pmatrix}\,,
\eal
where
\bal
\tilde{\cA}_{3/8}^0 =A_{\varphi} +\frac{i}{k \Pi^f} (r^1 \bar{r}^f -r^{f2} \bar{r}_2) +\tilde{M}_{\dot{b}}^{f\dot{a}} \tilde{r}_{\dot{a}}^{f} \bar{\tilde{r}}^{f\dot{b}}\,,
\eal
with
\bal
\tilde{M}_{\dot{b}}^{f\dot{a}}=\begin{pmatrix}
    0 & \frac{i}{k \tilde{\Lambda}} + \delta_{\dot{1}} \omega^f \bar{\beta}^{\perp}\\
    \frac{i}{k \tilde{\Lambda}} + \bar{\delta}^{\dot{1}} \Pi^f \beta_{\perp} & \bar{\beta}^{\parallel}
\end{pmatrix}\,.
\eal

Again generally the hyperloops obtained above are 1/16 BPS that are independent from \eqref{Pi=0loops1/4} and the $\Pi=0$ ones in \cite{Drukker:2022ywj}. Analogous to section \ref{Xi=0}, there are points of enhanced supersymmetry on the moduli space as well. Those loops with $\bar{\beta}^{\parallel}=0$ and $\delta_{\dot{1}} \omega^f \bar{\beta}^{\perp}= \bar{\delta}^{\dot{1}} \Pi^f \beta_{\perp}$ which is consistent with solutions of \eqref{sol3/8Pi=0} are enhanced to be 1/8 BPS\footnote{In the $\Pi^f\ne 0$ analogy, one might expect to find 3/16 BPS loops in this case as well, but it is in fact impossible because of the preserved $SU(2)_R$ R-symmetry.}, preserving two supercharges
\bal
\eta_a^{\imath} Q_{\imath}^{\dot{1}a} +\omega^{f\imath} \bar{u}_a (\sigma^1)_{\imath}^{\bar{\imath}} Q_{\bar{\imath}}^{\dot{1}a},\quad \eta_a^{\imath} Q_{\imath}^{\dot{2}a} +\omega^{f\imath} \bar{u}_a (\sigma^1)_{\imath}^{\bar{\imath}} Q_{\bar{\imath}}^{\dot{2}a}\,,
\eal
where $\omega^{f\imath}$ are components of $\omega^f$ with $\omega^f=\omega^{fl} v_l +\omega^{fr} v_r$, according to its definition in \eqref{omegaf}.

\section*{Acknowledgements}

We would like to thank N. Drukker for suggesting us to write this paper, for his support and inspirational comments throughout the project, and for his great help in improving the draft.
And we would like to acknowledge related collaboration with N. Drukker, M. Probst, M. Tenser and D. Trancanelli.
The work of ZK is supported by CSC grant No. 201906340174.
ZK would like to thank the University of Barcelona and the DESY theory group for their hospitality in the course of this work.

\appendix

\section{Supersymmetry transformation}
\label{susytrans}
As stated in the Introduction, we define the theory on $S^3$ and the hyperloops are located along the equator of this sphere. The supersymmetry transformations of twisted hypermultiplets are listed below, for the full details, see \cite{Drukker:2020dvr,Drukker:2022ywj}.
\bal
\label{SUSY2}
\delta \tilde q_{I-1\,\dot b}&=-\xi_{a\dot b}\tilde\psi_{I-1}^{a}\,,
\qquad 
\delta\bar{\tilde q}_{I-1}^{\,\dot b}=-\xi^{a\dot b}\bar{\tilde\psi}_{I-1\,a}\,,
\hskip8cm\\
\delta\tilde\psi_{I-1}^{a}&=-i\gamma^\mu\xi^{a\dot b}D_\mu \tilde q_{I-1\,\dot b}
-i\zeta^{a\dot b}\tilde q_{I-1\,\dot b}
+\frac{i}{k}\xi^{a\dot b}(\tilde q_{I-1\,\dot b}\tilde\nu_{I} -\tilde\nu_{I-1}\tilde q_{I-1\,\dot b})
\hskip-10cm
\\&\quad
-\frac{2i}{k}\xi^{b\dot c}\left(
\tilde q_{I-1\,\dot c\,}\mu_{I}{}_{\ b}^{a}
-\mu_{I-1}{}_{b}^{\ a}\tilde q_{I-1\,\dot c}\right),
\hskip-1cm
\\
\delta\bar{\tilde\psi}_{I-1\,a}&=-i\gamma^\mu\xi_{a\dot b}D_\mu \bar{\tilde q}_{I-1}^{\,\dot b}
-i\zeta_{a\dot b}\bar{\tilde q}_{I-1}^{\,\dot b}+\frac{i}{k}\xi_{a\dot b}(\tilde\nu_I \bar{\tilde q}_{I-1}^{\,\dot b}-\bar{\tilde q}_{I-1}^{\,\dot b}\tilde\nu_{I-1})
\hskip-10cm
\\&\quad
-\frac{2i}{k}\xi_{b\dot c}\left(
\mu_{I}{}_{\ a}^{b}\bar{\tilde q}_{I-1}^{\,\dot c}
-\bar{\tilde q}_{I-1}^{\,\dot c}\mu_{I-1}{}_{a}^{\ b}\right),
\eal
where $\xi^{a\dot b}$ is a linear combination of four Killing spinors $\xi^l$, $\xi^{\bar l}$, $\xi^r$, $\xi^{\bar r}$ on $S^3$
\beq
\xi_{\alpha}^{a\dot b}=\xi_{\imath}^{a\dot b}\xi^\imath_\alpha + \xi_{\bar\imath}^{a\dot b}\xi^{\bar\imath}_\alpha\,,
\eeq
where $\imath = l, r$, $\bar{\imath}=\bar{l}, \bar{r}$, and the spinor index $\alpha=\pm$. Along the circle we take
$\gamma_\varphi=\sigma_3$ and the Killing spinors reduce to 
\cite{Assel:2015oxa}
\beq
\label{killingspinors}
\xi^l_\alpha=\begin{pmatrix}1\\0\end{pmatrix},
\qquad
\xi^{\bar l}_\alpha=\begin{pmatrix}0\\1\end{pmatrix},
\qquad
\xi^r_\alpha=\begin{pmatrix}e^{-i\varphi}\\0\end{pmatrix},
\qquad
\xi^{\bar r}_\alpha=\begin{pmatrix}0\\e^{i\varphi}\end{pmatrix},
\eeq
whence one finds $\zeta^{l,\bar l}_{a\dot b}=\frac{i}{2}\xi^{l,\bar l}_{a\dot b}$ and 
$\zeta^{r,\bar r}_{a\dot b}=-\frac{i}{2}\xi^{r,\bar r}_{a\dot b}$.

\bibliographystyle{utphys2}
\bibliography{refs}
\end{document}